\documentclass[aps, a4paper, twocolumn, nofootinbib, showpacs, amsfonts]{revtex4}

\usepackage[utf8]{inputenc}
\usepackage{ dsfont }
\usepackage{amssymb}

\usepackage{graphicx}
\usepackage{amsmath}
\usepackage{bm}
\usepackage{xcolor}
\usepackage{tikz}
\usepackage{mathrsfs}
\usetikzlibrary{arrows}
\usepackage{cancel}

\newcommand{\bea}{\begin{eqnarray}}
\newcommand{\eea}{\end{eqnarray}}
\newcommand{\be}{\begin{equation}}
\newcommand{\ee}{\end{equation}}

\def\nn{\nonumber}

\begin{document}

\title{Observational properties of Coherent Quantum Black Holes}

\author{Abdybek Urmanov}
\email{abdybek.urmanov@nu.edu.kz}
\affiliation{Department of Physics, Nazarbayev University, Kabanbay Batyr 53, 010000 Astana, Kazakhstan}

\author{Hrishikesh Chakrabarty}
\email{hrishikesh.chakrabarty@nu.edu.kz}
\affiliation{Department of Physics, Nazarbayev University, Kabanbay Batyr 53, 010000 Astana, Kazakhstan}

\author{Daniele Malafarina}
\email{daniele.malafarina@nu.edu.kz}
\affiliation{Department of Physics, Nazarbayev University, Kabanbay Batyr 53, 010000 Astana, Kazakhstan}
    
\begin{abstract}
We consider null and time-like geodesics around a spherically symmetric, non-rotating Coherent Quantum Black Hole (CQBH). 
The classical limit of the geometry of CQBH departs from that of the Schwarzschild spacetime at short scales and depends on one parameter $R_s$ which can be interpreted as the physical radius of the `quantum' core. We study 
circular orbits, photon rings, and lensing effects 
and compare them with the Schwarzschild metric. Using the relativistic ray-tracing code \texttt{GYOTO}, we produce a simulation of the shadow and show that thin accretion disks around a CQBH possess unique ring structures that distinguish them from other theoretical models. 

\end{abstract}

\maketitle

\section{Introduction}\label{sec1}

The term `black hole mimickers' refers to theoretical astrophysical objects that closely resemble black holes in observational appearance, exhibiting similar features such as photon rings and accretion disks while at the same time being described by a theoretical model distinct from classical black holes \cite{Cardoso:2019rvt}.

Various types of black hole mimickers have been proposed over the years as candidates for the extreme compact objects we observe in the universe. These include gravastars \cite{Cardoso:2016oxy,Visser:2003ge,Chirenti:2007mk,Harko:2009gc,Broderick:2007ek}, boson stars \cite{Vincent:2015xta,Olivares:2018abq,Marunovic:2013eka,Rosa:2023qcv}, wormholes \cite{Einstein:1935tc,Berry_2020,Berry:2020ntz,PhysRevD.105.023021, PhysRevD.101.104021}, regular black holes \cite{Casadio:2024lgw,PhysRevLett.96.031103,bardeen2014black,bardeen2019models,Bambi_2013,PhysRevD.105.064065,Mazza_2021,Abdujabbarov:2016hnw,Lemos:2008cv,PhysRevLett.132.031401} and deformed compact objects \cite{Carballo-Rubio:2018jzw,Shaikh_2022,Narzilloev_2020,Abdikamalov:2019ztb,Chowdhury:2011aa}. 
In recent times, significant progress in experimental observations, such as the imaging of the shadow of the supermassive black hole at the core of the M87 galaxy \cite{EventHorizonTelescope:2019dse, EventHorizonTelescope:2019ths} and the detection of gravitational waves from binary mergers \cite{PhysRevLett.116.061102} has presented the opportunity to test the validity of these alternative models and in turn potentially constrain proposed modification to General Relativity (GR) \cite{Li:2022eue, Shaikh_2023, Johnson_McDaniel_2020, PhysRevLett.116.171101}.

In the present article we consider an alternative black hole model, called Coherent Quantum Black Hole (CQBH), which is obtained from a simple prescription for quantization that was proposed by Casadio \cite{Casadio:2021eio, Casadio:2021onj,Casadio:2021cbv}. 
The CQBH, despite being a simple toy model, presents several features that make it appealing as a hypothetical candidate for an astrophysical compact object. First of all it is built with a minimal approach for the quantization of gravity in the vicinity of the classical black hole singularity which 
produces a compact massive core of finite size. Secondly, the core of the resulting object may be located below the classical event horizon. This results in the CQBH having an outer horizon which mimics the black hole event horizon while lacking an inner Cauchy horizon, thus avoiding the problems related to causality violations and mass inflation \cite{Price:1971fb, Simpson:1973ua,Poisson:1989zz, Poisson:1990eh,Carballo-Rubio:2018pmi}.
Finally, the CQBH produces a spacetime geometry which bears the unique signature of its core. This means that quantum corrections in the strong curvature region produce macroscopic effects which are potentially measurable. From the point of view of observations this may allow for observational tests to constraint the size of the object's core and consequently the scale of the modifications to GR.
In \cite{Casadio:2023pmh} it was shown that the CQBH is consistent with the Bekenstein-Hawking thermodynamic entropy for black holes
while in \cite{Casadio:2022ndh} a charged CQBH was obtained.
In this article we study the motion of massive test particles and photons in the geometry of the CQBH and investigate how future observations may be used to test the physical viability of this hypothetical compact object.

The paper is organized as follows: In section \ref{sec2} we briefly review the line element and its properties. Section \ref{sec3} and \ref{sec4} are dedicated to modelling the motion of massive and massless particles respectively. In particular in section \ref{sec3} we model thin accretion disks in the geometry of the CQBH while in section \ref{sec4} we model gravitational lensing. In section \ref{sec5} we use \texttt{GYOTO}, a general relativistic raytracing code \cite{gyoto,Aimar:2023vcs,Paumard_fvincent_Gourgoulhon_LamyFrederic_Torrance_iurso_2024} to simulate the image of the CQBH's shadow. Finally in section \ref{sec6} we summarize the results and their implications.
Throughout the article we use geometrized units setting $c=1$ with the metric signature $\{-,+,+,+\}$.

\section{The spacetime}\label{sec2}

\begin{figure}[tt]
   \begin{center}
        \includegraphics[width=9cm]{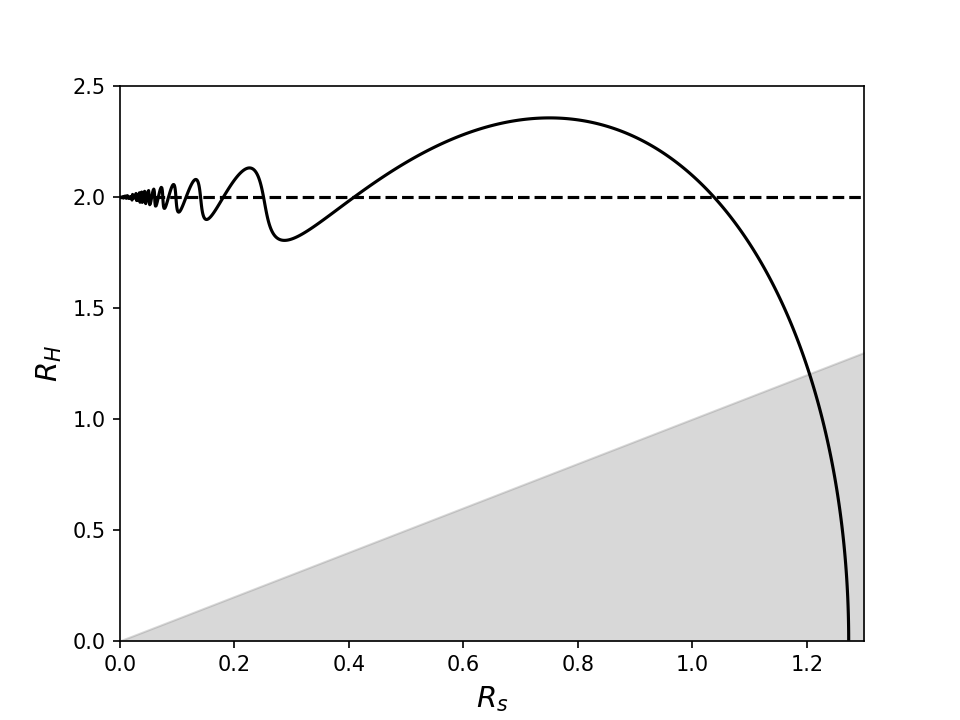}    
  \end{center}
    \caption{ The solid line represents the horizon radius $R_H$ of the CQBH as a function of $R_s$. For comparison we show the Schwarzschild radius $R_{\rm Sch}$ given by the horizontal dashed line. The shaded area represents the size of the matter core showing that for $R_s \gtrsim 1.2$ the core is larger than the horizon. }\label{fig:RH}
\end{figure}

The geometry of the static and spherically symmetric CQBH in Schwarzschild coordinates 
is given by 
\begin{equation}
    ds^2=-(1+2V_{\rm QN})dt^2+\frac{dr^2}{1+2V_{\rm QN}}+r^2d\Omega^2,
    \label{metric}
\end{equation}
where $ d\Omega^2 $ is the line element on the unit sphere and
\begin{equation}
    V_{\rm QN}=\frac{2V_{\rm N}}{\pi}{\rm Si}\left(\frac{r}{R_s}\right),
    \label{Vqn}
\end{equation}
represents the effective quantum analog of the Newtonian potential $V_{\rm N}=-G_{\rm N}M/r$. Here $R_s$ is a parameter describing the size of the quantum core 
and 
\be
{\rm Si}(x)=\int_{0}^{x}\frac{\sin{z}}{z}dz,
\ee
is the Sine integral function. The other constants are $G_{\rm N}$, i.e. Newton's constant and $M$, which describes the ADM mass of the CQBH. 
It is important to notice that the parameter $R_s$ in the CQBH model is obtained by introducing an Ultra-Violet (UV) cutoff scale to regularize the UV divergence due to the vanishing volume and infinite density of the classical source of the Schwarzschild metric. 
Therefore the metric must be understood as a simple toy model that captures the essence of the classical configuration emerging from the coherent quantum states.
The resulting core is `classical' in the sense that it is larger than the Planck length and of the order of the Schwarzschild radius \cite{Casadio:2021onj}. 
In fact $R_s$ may be larger, smaller or equal than the size of the horizon radius (see Fig.~\ref{fig:RH}). 
Due to this additional parameter in the metric, the solution violates the no-hair theorem \cite{Calmet:2021stu}.

The CQBH metric \eqref{metric} reduces to the Schwarzschild metric in the weak field limit, as can be seen by taking the limit, as $ r/R_s \rightarrow \infty $, for which the sine integral function $ {\rm Si}\left(r/R_s\right) \rightarrow \pi/2 $ and $ V_{\rm QN} \simeq V_{\rm N} $. Hence, at large distances the metric is asymptotically flat.      
However at distances close to the core, the properties of the CQBH differ significantly from those of a Schwarzschild black hole.
For instance the classical black hole singularity is replaced by a finite size core with
an integrable singularity at the center and finite tidal forces \cite{Casadio:2021eio, intsing}. 
This suggests that observations of orbits in the vicinity of the compact object may allow to distinguish it from a Schwarzschild black hole.

The radius of the event horizon $R_H$ of a CQBH is obtained by solving the equation $ g^{rr} = 1+2V_{\rm QN} = 0$ and is shown in Fig.~\ref{fig:RH} as a function of the core radius $R_s$. 
Notice that due to the presence of the Sine integral function, the value of the horizon radius as a function of $R_s$ oscillates about 
the Schwarzschild radius ($ R_{\rm Sch} = 2M $). 
In the following, for the sake of simplicity, we take the ADM mass of the compact object as $M=1$, or equivalently we rescale the time and radial coordinates $t$, $r$ and the core radius $R_s$ in units of $M$. 
We can see that there exist values of $R_s$ for which the horizon of the CQBH is the same as Schwarzschild. The largest value of $R_s$ for which the two horizons coincide is $ R_s \simeq 1.06 $ and for $ R_s \gtrsim 1.2 $ the horizon radius is smaller than the core radius, thus making the CQBH a horizonless compact object.
A detailed analysis of the CQBH can be found in \cite{Casadio_Micu_2024}. In the following, our focus lies in understanding the properties of massive and massless particle motion in the CQBH exterior and how specific features of lensing, shadow and accretion disks depend on the parameter $R_s$.

\section{Motion of massive particles}\label{sec3}

Black holes are not directly observable by definition. However, their surroundings have properties that depend directly on the nature and properties of the central object. For example, the closest stable orbit to a Kerr black hole depends on the value of the central object's angular momentum. Therefore, at least in principle, it should be possible to test theoretical models of extreme compact objects through observations of their surrounding environment, such as accretion disks.
We can model the motion of particles in the accretion disk by considering geodesic motion for test particles (assumed to have negligible mass as compared to the central object's) on circular orbits. To this aim in the following we shall write the equations of motion for test particles.

Due to the static and spherically symmetric nature of the CQBH metric, 
there exist two Killing vectors corresponding to two constants of motion for particles on geodesics, namely the particle's energy $p_t = \Tilde{E}$ and angular momentum $p_\phi = \Tilde{L}$.
Also, due to spherical symmetry, we can restrict the analyses to the equatorial plane. 
Now, the conserved quantities per unit mass $E=\Tilde{E}/m$ and $L=\Tilde{L}/m$ are obtained from 
\bea
    \dot{t}&=&\frac{p^t}{m}
    =(1+2V_{\rm QN})^{-1}E,
    \label{mass_dt} \\
    \dot{\phi}&=&\frac{p^{\phi}}{m}
    =r^{-2}L,
    \label{mass_dp}
\eea
where $\dot{x}^\mu=dx^\mu/d\tau$ is the tangent to the geodesic $\{x^\mu(\tau)\}$ with $\tau$ the proper time. Using the normalization $g_{\mu \nu}\dot{x}^{\mu} \dot{x}^{\nu}=-1$ for massive particles, we find the radial equation of motion as
\begin{equation}
    \dot{r}^2=E^2-V_{\rm eff}(r),
    \label{mass_dr}
\end{equation}
where $V_{\rm eff}$ is the effective potential which for the metric \eqref{metric} is given by
\begin{equation}
        V_{\rm eff}=\left(\frac{L^2}{r^2}+1\right)\left(1+2V_{\rm QN}\right).
        \label{veff1}
\end{equation}

\begin{figure}
  \includegraphics[width=0.5\textwidth]{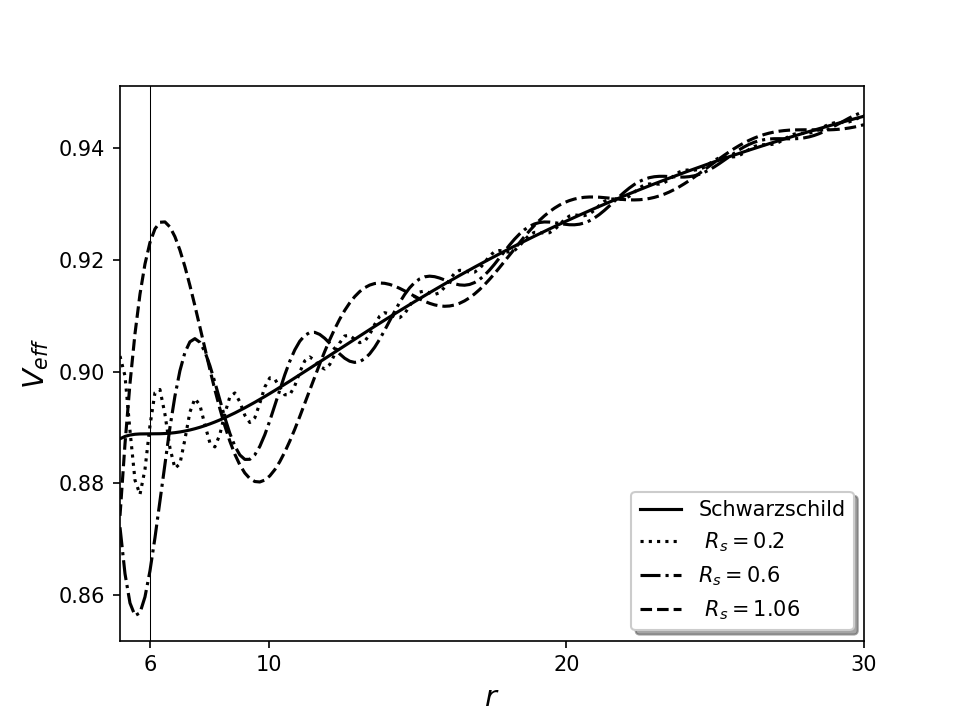}
  \caption{ The effective potential $V_{\rm eff}$ as a function of the radial coordinate for massive particles in the equatorial plane of CQBH spacetime as compared to the Schwarzschild effective potential (solid line). All cases 
  are shown for angular momentum $L=12$. The vertical grey line highlights the innermost stable circular orbit for the Schwarzschild case.}
  \label{fig:Veff}
\end{figure}

In Fig.~\ref{fig:Veff}, we plot the effective potential 
with respect to the radial coordinate $r$ 
and compare its behavior with that of a Schwarzschild black hole. It is interesting to note that the effective potential for massive particles around a CQBH exhibits an oscillatory behavior with several local extrema for any given values of $E$ and $L$, which is in sharp contrast with the Schwarzschild case. In Fig.~\ref{fig:Veff}, we chose the value of the angular momentum $L=12$ which corresponds to particles on circular orbit rotating at the innermost stable circular orbit (ISCO), $R_{\rm ISCO}=6$, in the Schwarzschild spacetime. 
The most notable feature is that there exist multiple radii allowing stable orbits even beyond the ISCO radius for Schwarzschild.
These numerous stable and unstable orbits could substantially affect the appearance of the accretion disk around the CQBH.
For larger values of $r$, \textit{i.e.} far from the black hole, the CQBH effective potential tends toward Schwarzschild and the potential barrier for stable orbits becomes smaller, thus making it more difficult for particle to circularize in a realistic accretion disk.

For timelike geodesics, the effective potential $V_{\rm eff}$ is given by Eq.~\eqref{veff1} and the energy and angular momentum of a particle on a circular orbit at a radius $r$ can be found from the conditions $ V_{\rm eff} = E$ and  $V'_{\rm eff} = 0 $, where the prime denotes derivative with respect to the radial coordinate $r$. More explicitly, the second equality gives 
\begin{equation}\label{eq:vprime}
        {V'_{\rm eff}}= \frac{2L^2}{r^3} + \frac{4L^2V_{\rm QN}}{r^3} - \frac{2L^2V'_{\rm QN}}{r^2} - 2V'_{\rm QN} = 0, 
\end{equation}
and using Eqs.~\eqref{mass_dr} and \eqref{eq:vprime}, we can obtain the energy and angular momentum $E(r)$ and $L(r)$ of particles on a given circular orbit with radius $r$ as 
\bea
    E^2&=&\frac{(1+2V_{\rm QN})^2}{1+2V_{\rm QN}-V'_{\rm QN}r}, \\
    L^2&=& \frac{r^3 V'_{\rm QN}}{1+2V_{\rm QN}-V'_{\rm QN}r}.
\eea
Similarly, the particle's angular velocity is given by 
\begin{equation}
    \omega = \frac{d\phi}{dt}=\sqrt{\frac{V'_{\rm QN}}{r}}.
\end{equation}
At distances far from the compact object, $V_{\rm QN} \simeq V_{\rm N} $, and these quantities reduce to those of the Schwarzschild metric.
Now, in Schwarzschild we can find the radius of the ISCO from the condition $V_{\rm eff}'' = 0$, which gives the radius at which the circular orbit is marginally stable. In the case of the CQBH the condition $V_{\rm eff}'' = 0$ identifies the radii at which the local extrema become marginally stable. We get
\bea \nn 
    {V_{\rm eff}}''&=&\frac{6L^2}{r^4}-\frac{8L^2V'_{\rm QN}}{r^3}+\frac{12L^2V_{\rm QN}}{r^4}+\\ 
    &&+\frac{2L^2V''_{\rm QN}}{r^2}+2V''_{\rm QN}= 0,
\eea
where 
\bea
    V_{\rm QN}'&=&\frac{2}{\pi}\frac{GM}{r^2} \left[{\rm Si}\left(\frac{r}{R_s}\right)-\sin{\frac{r}{R_s}}\right], \\
    V_{\rm QN}''&=& \frac{2}{\pi}\frac{GM}{r^2}\left[-\frac{2}{r}{\rm Si}\left(\frac{r}{R_s}\right)+\frac{3}{r}\sin{\frac{r}{R_s}}-\frac{1}{R_s}\cos{\frac{r}{R_s}} \right].\notag\\
\eea

\begin{figure}
  \includegraphics[width=0.5\textwidth]{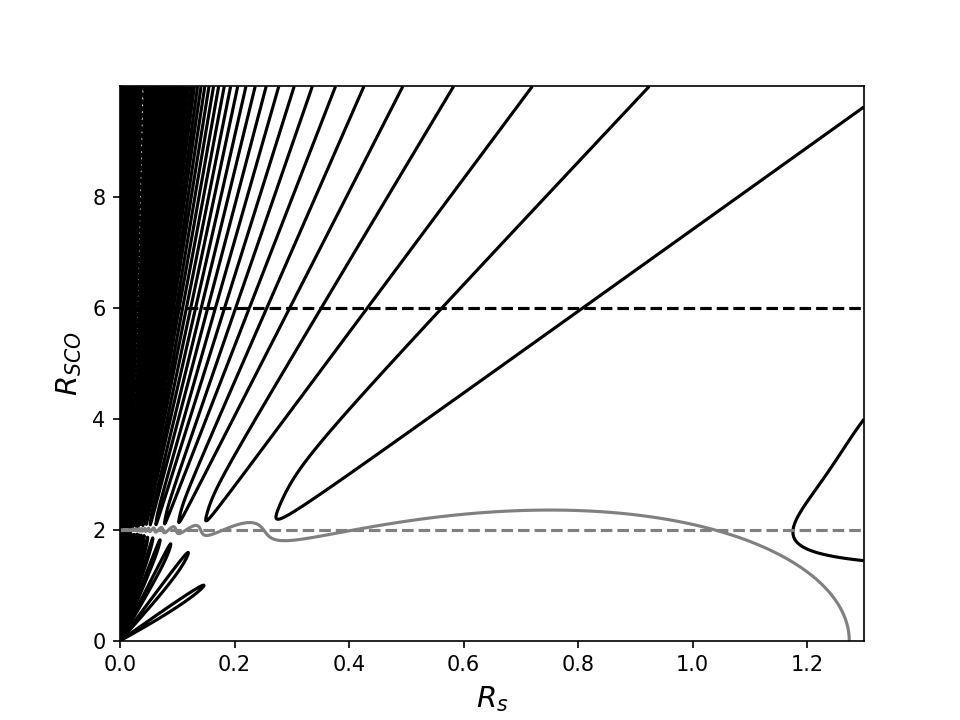}
  \caption{ Boundary contours for regions of stable circular orbits for the CQBH depending on the value of the parameter $R_s$. The black horizontal dashed line represents $R_{\rm ISCO}$ for the Schwarzschild case. The grey solid and dashed lines denote the radius of the horizons for the CQBH and Schwarzschild, respectively.}
  \label{fig:risco}
\end{figure}

Stable circular orbits for massive particles around the CQBH are shown in Fig.~\ref{fig:risco}, depending on the radius of the core $R_s$. 
For any given value of $R_s$, there exist several regions of stable circular orbits in the CQBH spacetime and the radius of the ISCO $R_{\rm ISCO}$ for each region corresponds to the smallest stable circular orbit. 
This means that there exist gaps between separate regions of stable orbits. This feature persists also at large distances making the CQBH disk's structure unique.
Of course, at large distances the gaps would become less prominent as the potential barriers become smaller.
Note that the smallest $R_{\rm ISCO}$ for the CQBH that resembles Schwarzschild is obtained for $ R_s \approx 0.8$.
Notice also that for values of $R_s \gtrsim 0.8$ the smallest ISCO radius is larger than Schwarzschild's.

On the other hand, the regions of stable circular orbits get closer for smaller values of $R_s$ and it can be expected that for sufficiently small values there would be stable orbits all the way
to the CQBH horizon. 
This is in sharp contrast to the Schwarzschild case where stable orbits can no longer exist beyond $r = 6$.

\section{Motion of massless particles}\label{sec4}

\begin{figure}
  \centering
  \includegraphics[width=0.5\textwidth]{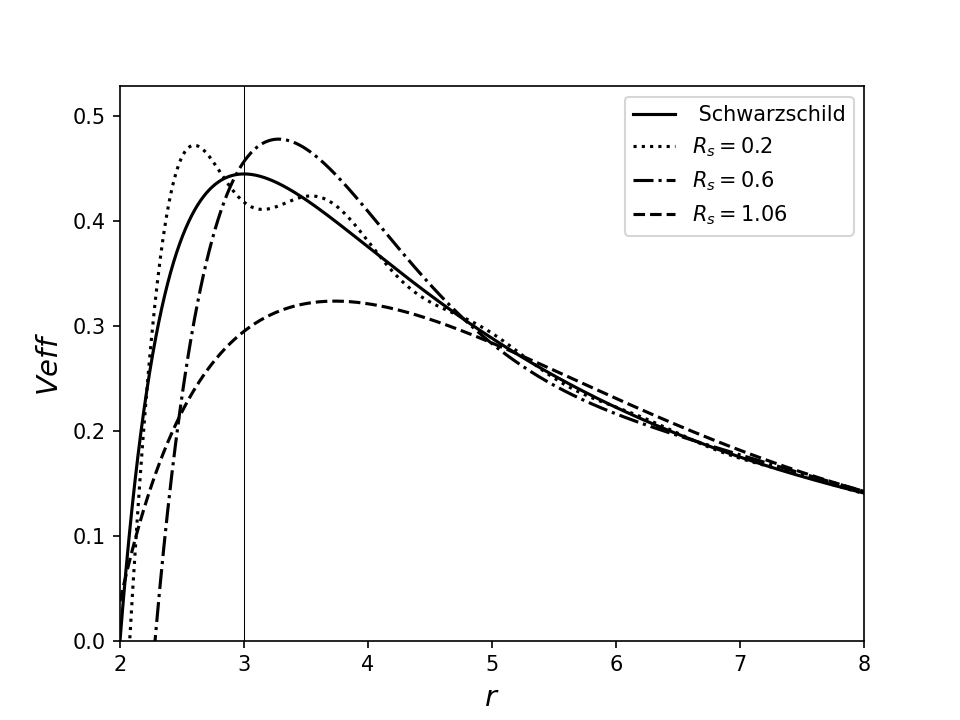}
  \caption{The effective potential for massless particles in the equatorial plane of the CQBH as a function of the radial coordinate in units of $M$. The effective potential is shown for three values of $R_s$. Notice that for $R_s=0.2$ (dotted line) there exist three locations for circular orbits for photons, with one of them being stable. For comparison the solid line shows the corresponding effective potential for Schwarzschild and the vertical grey line highlights the Schwarzschild photon ring radius.}
  \label{fig:Veffph}
\end{figure}

Following the same procedure as before for massless particles, we still have the conserved quantities from Eqs.~\eqref{mass_dt} and \eqref{mass_dp} and we must now use
$g_{\mu \nu}\dot{x}^{\mu} \dot{x}^{\nu}=0$ to write the radial equation of motion
\begin{equation}
    \dot{r}^2=E^2-\frac{L^2}{r^2}(1+2V_{\rm QN}),
    \label{photons_dr}
\end{equation}
so that the effective potential for photons is given by
\begin{equation}
    V_{\rm eff}=(1+2V_{\rm QN})\frac{L^2}{r^2}.
\end{equation} 
The photon ring is defined as the circular orbit of massless particles and it is given by the condition $V'_{\rm eff} = 0$, explicitly 
\begin{equation}
1-V_{\rm QN}'r+2V_{\rm QN}=0.
\end{equation}
This equation reduces to 
$$
\frac{\pi r}{2G_NM}=3{\rm Si}\left(\frac{r}{R_s}\right)-\sin \left(\frac{r}{R_s}\right),
$$ 
which can be solved numerically to obtain the radius of photon sphere. The most interesting aspect is that for $R_s \lesssim 0.36$ there exist values of $R_s$ for which the above equation has multiple solutions, indicating the existence of unstable as well as stable orbit for photons. In Fig.~\ref{fig:Veffph}, we compare the effective potentials for massless particles in Schwarzschild and CQBH for different values of $R_s$. Again the effective potential of the CQBH shows an oscillatory behavior about the Schwarzschild's effective potential with the most interesting feature being the existence of one stable circular orbit for photons for $R_s\simeq 0.2$.

\begin{figure}
  \centering
  \includegraphics[width=0.5\textwidth]{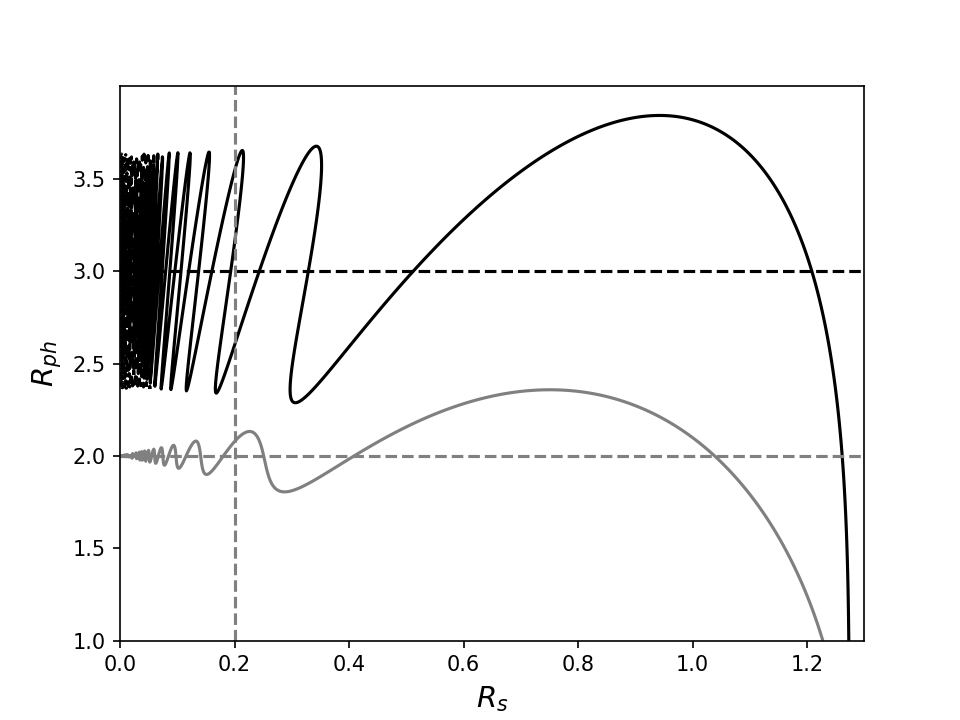}
  \caption{The CQBH's photon ring radius ($R_{\rm ph}$) as a function of the parameter $R_s$ is represented by the solid black line in comparison with Schwarzschild's photon ring as the dashed black line. Also, the event horizon for the CQBH and Schwarzschild are shown by the grey solid and dashed lines, respectively. The vertical line shows the location of the three circular photon orbits obtained for $R_s=0.2$ as can be seen by the extrema of the effective potential in Fig.~\ref{fig:Veffph}.}
  \label{fig:Rph}
\end{figure}

In Fig.~\ref{fig:Rph}, we show the radius of the CQBH photon ring as a function of parameter $R_s$. For $R_s =1.06$, the horizon of CQBH is approximately equal to the Schwarzschild one. However, at this value of the parameter, the photon ring of the CQBH is larger than the Schwarzschild photon ring. Also, for smaller $R_s$ there are regions where more than one photon ring may appear, for example, there are three rings for $R_s \simeq 0.2$, and for even smaller values of the parameter the photon rings get closer. At sufficiently small values of $R_s$ we expect the existence of a photon disk which spreads from $r\approx2.35$ to $r\approx3.6$.

\subsection{Lensing}  
The first application of General Relativity was gravitational lensing \cite{1936Sci....84..506E}
and since then it has been widely used to study several astrophysical phenomena (see, for example~\cite{1992grle.book.....S,luminet,Clowe:2006eq,Virbhadra:1999nm,Abdujabbarov:2017pfw}).
In the following, we analyze the deflection angle of light in the CQBH geometry and compare it with that by a Schwarzschild black hole. 

Using Eq.~\eqref{mass_dp} and \eqref{photons_dr}, we obtain 
\begin{equation}
    \frac{d\phi}{dr}=\pm\frac{1}{r^2}\Bigg[\frac{1}{b^2}-\frac{1}{r^2}(1+2V_{\rm QN})\Bigg]^{-1/2},
\end{equation}
where $b=L/E$ is the impact parameter for a massless particle. Then the magnitude of the total angle swept by a light-ray as it travels from infinity towards the closest approach to the CQBH and to infinity again can be obtained by simply integrating the previous equation 
\begin{equation}
 \Delta\phi=2\int_{r_1}^{\infty}\frac{dr}{r^2}\Bigg[\frac{1}{b^2}-\frac{1}{r^2}(1+2V_{\rm QN})\Bigg]^{-1/2},
\end{equation}
where due to symmetry we can integrate from infinity to the turning point $r_{1}$, defined by the condition $dr/dt\big|_{r_1}=0$ and for which $V_{\rm eff}(r_1)=1/b^2$. Introducing a new variable $w=b/r$ we rewrite $\Delta\phi$ as 
\begin{equation}
    \Delta\phi=2\int_{0}^{w_1}dw\left[1-w^2\left(1-\frac{4Mw}{\pi b}{\rm Si}\left(b/wR_s\right)\right)\right]^{-1/2},
\end{equation}
where $w_1=b/r_1$ and we have set $G_{\rm N}=1$. 

\begin{figure}[hh]
  \centering
  \includegraphics[width=0.5\textwidth]{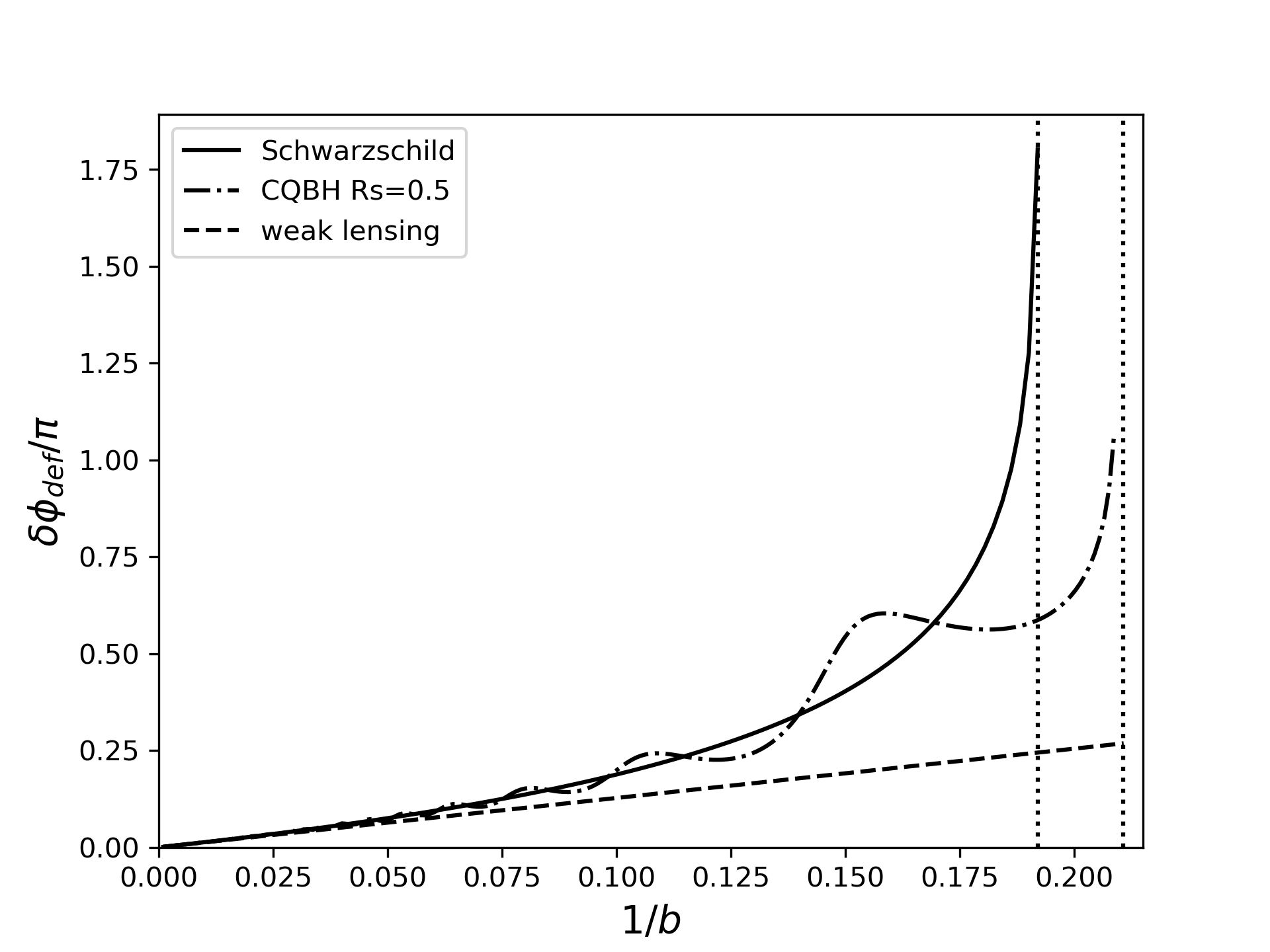}
  \caption{The deflection angle $\delta\phi$ as a function of the inverse of the impact parameter $1/b$. The solid line shows the Schwarzschild case while the dash-dotted line shows the CQBH with $R_s=0.5$. The dashed line represents the weak lensing limit. The two dotted vertical lines denote the critical value of impact parameter $b_{\rm crit}$ at which the photons get captured.  }
  \label{fig:defang}
\end{figure}

Of course, it is not possible to find an analytical solution for the integral above, because of the presence of the sine integral function. Therefore, first we look at the weak-field limit, i.e when $M/b\ll 1$.
In the weak-field limit, as $b/wR_s\rightarrow\infty$ and for large values of the argument the Sine integral function ${\rm Si}(x)$ can be rewritten as an asymptotic series as

\be \label{LargeSi}
    {\rm Si}(x)  \simeq \frac{\pi}{2} - \frac{\cos{x}}{x} 
    - \frac{\sin{x}}{x^2}  + \cdots
\ee
Keeping the next to leading order term from Eq.~\eqref{LargeSi} 
$\Delta\phi$ becomes
\begin{equation} \label{deltaphi}
    \Delta\phi=2\int_{0}^{w_1}dw\left[1-w^2\left(1-\frac{2M}{b}w+H(w)\right)\right]^{-1/2},
\end{equation}
with
\be 
H(w)=\frac{4MR_s}{\pi b^2}w^2\cos\left(\frac{b}{wR_s}\right).
\ee 
Notice that $H(w)$ has a term in $MR_s/b^2$ which in the weak field limit is negligible. Then for $H=0$ Eq.~\eqref{deltaphi} is same as in the Schwarzschild case and can be integrated to find $\Delta\phi=\pi+4M/b$. Considering $H\neq 0$ and making an expansion of the integrand in Eq.~\eqref{deltaphi} for $M/b \ll 1$ we get    
\begin{equation}
    \Delta\phi=\pi+\frac{4M}{b}-\frac{4MR_s}{\pi b^2}\int_{0}^{w_1}dw \frac{w^4 \cos(b/wR_s)}{(1-w^2)^{3/2}},
\end{equation}
which is related to the deflection angle for a light ray via $\delta \phi = \Delta\phi - \pi$.
The deflection angle as a function of the inverse of the impact parameter $1/b$ for Schwarzschild (with $M=1$) and for the CQBH with $R_s=0.5$ is shown in Fig.~\ref{fig:defang}. 

We can also calculate the critical value of the impact parameter $b_{\rm crit}$ at which photons get captured into the photon ring. The critical impact parameter can be calculated numerically by substituting $R_{\rm ph}$ into Eq.~\eqref{photons_dr} and solving the following equation
\begin{equation}
    \frac{1}{b_{\rm crit}^2}-\frac{1+2V_{QN}(R_{\rm ph})}{R^2_{\rm ph}}=0.
\end{equation}
The value of $b_{\rm crit}$ is important because it determines the radius of the shadow of the black hole which in turn relates to the image seen by a distant observer. Then for Schwarzschild we know that $b_{\rm crit}=3\sqrt{3}\approx5.196$ and from Fig.~\ref{fig:defang} for a CQBH with $R_s=0.5$ we find $b_{\rm crit}\approx4.764$.
Fig.~\ref{fig:defang} also shows that lensing for CQBH and Schwarzschild coincides in the weak-field regime. However, for any given value of $R_s$ the value of the deflection angle of the CQBH as a function of $1/b$ oscillates about the Schwarzschild value in a manner that may
affect the appearance of the accretion disk and the shadow. 

\section{Shadow}\label{sec5}

\begin{figure*}
    \centering
    \begin{tabular}{llll}
        $\theta=5^\circ $ & 
        \includegraphics[width=.3\linewidth]{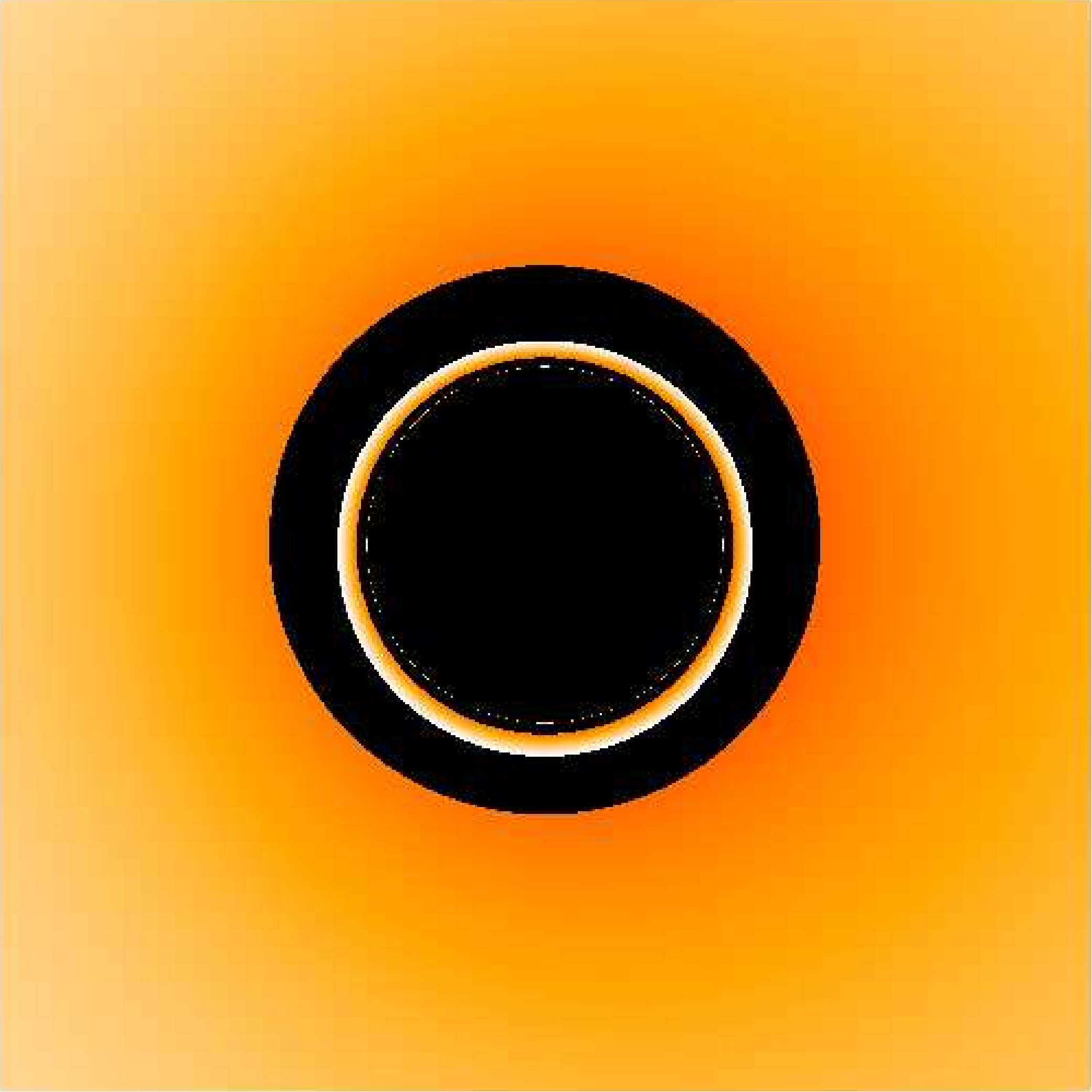} & \includegraphics[width=.3\linewidth]{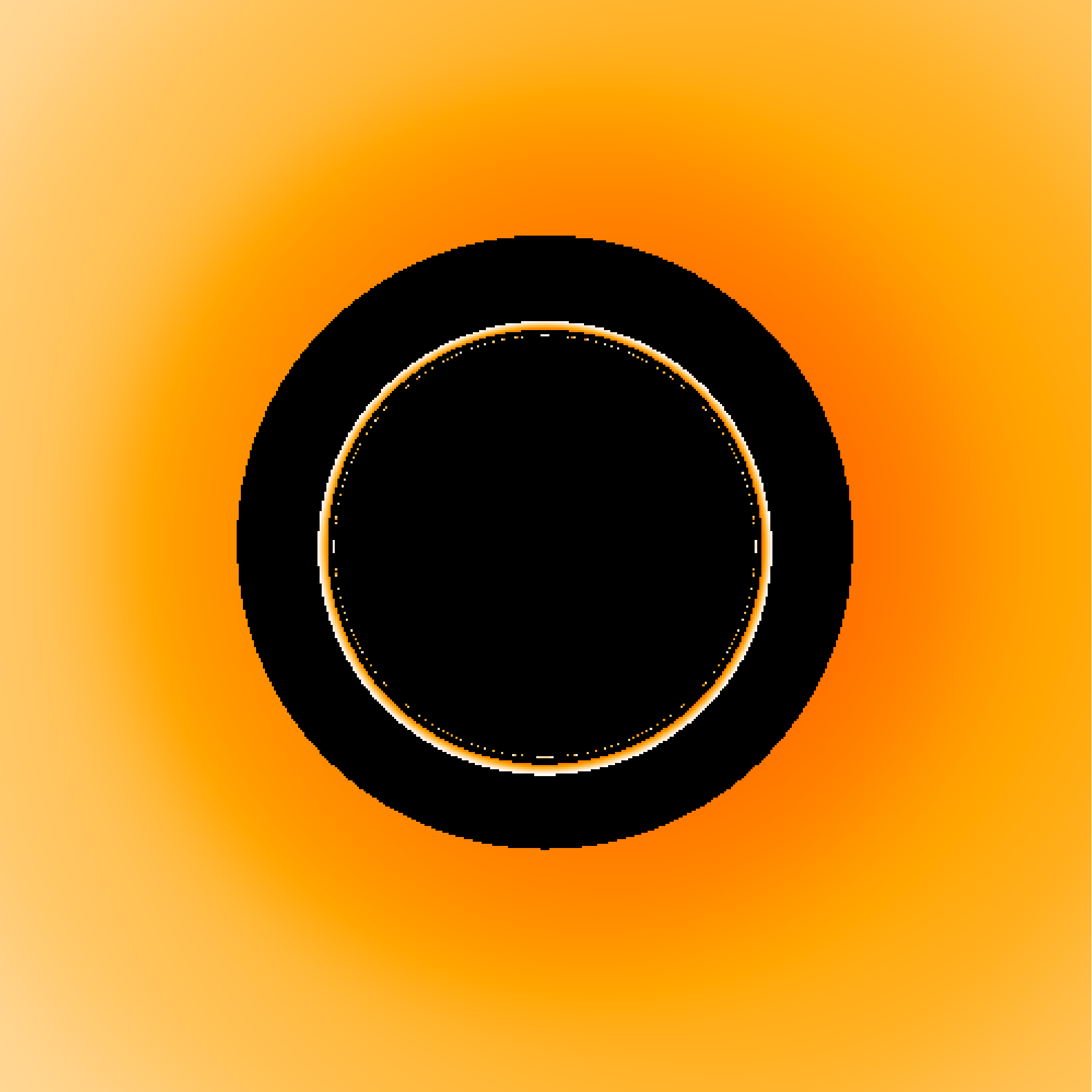} & \includegraphics[width=.3\linewidth]{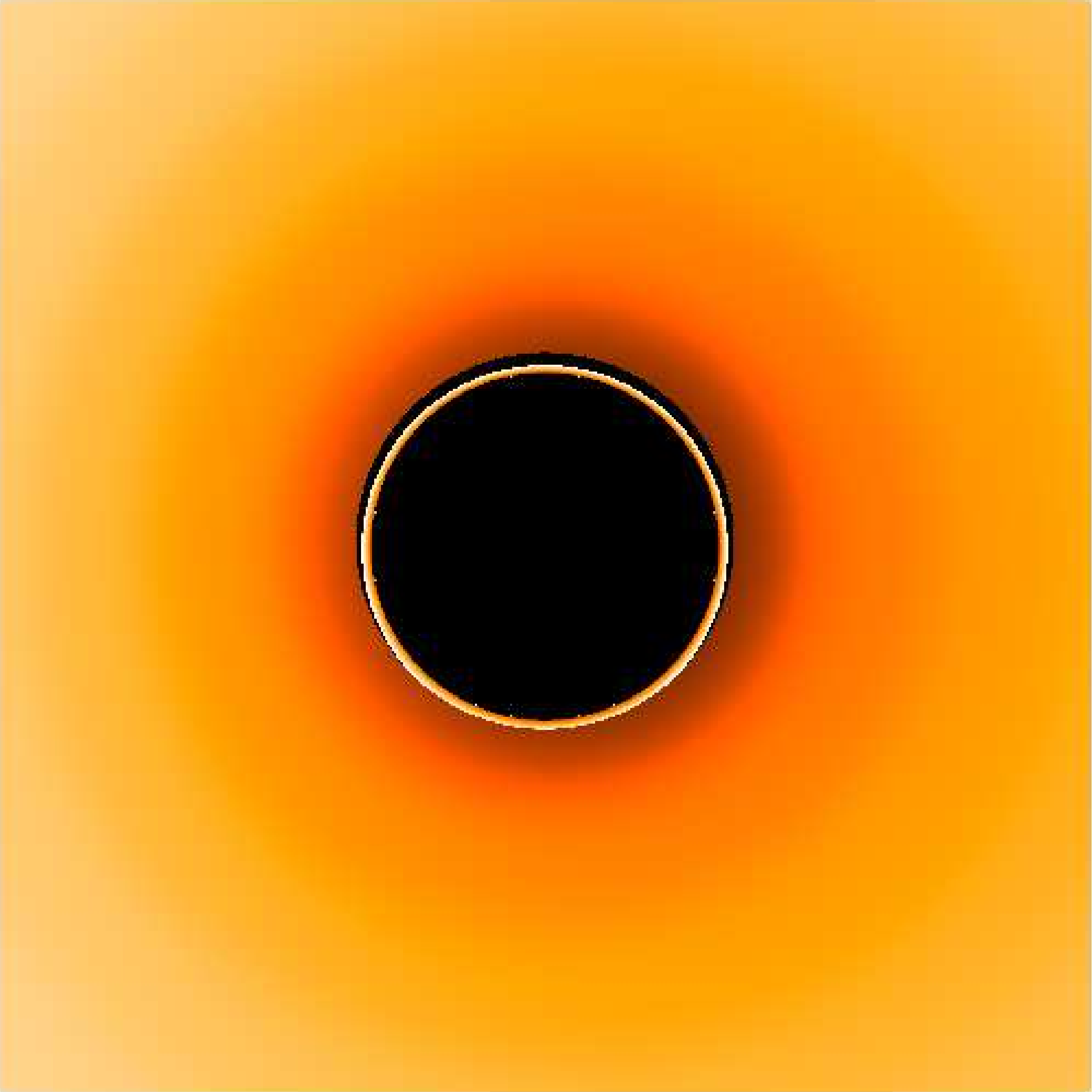}\\
        $\theta=72^\circ$ & 
        \includegraphics[width=.3\linewidth]{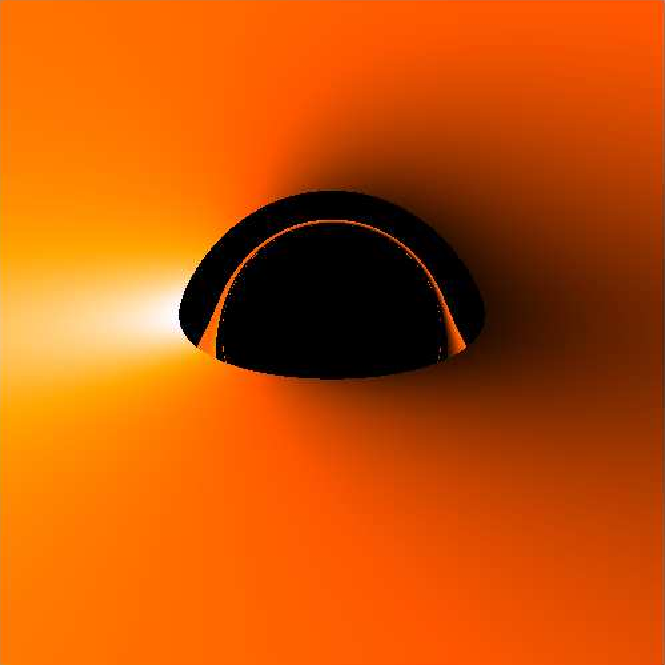} & \includegraphics[width=.3\linewidth]{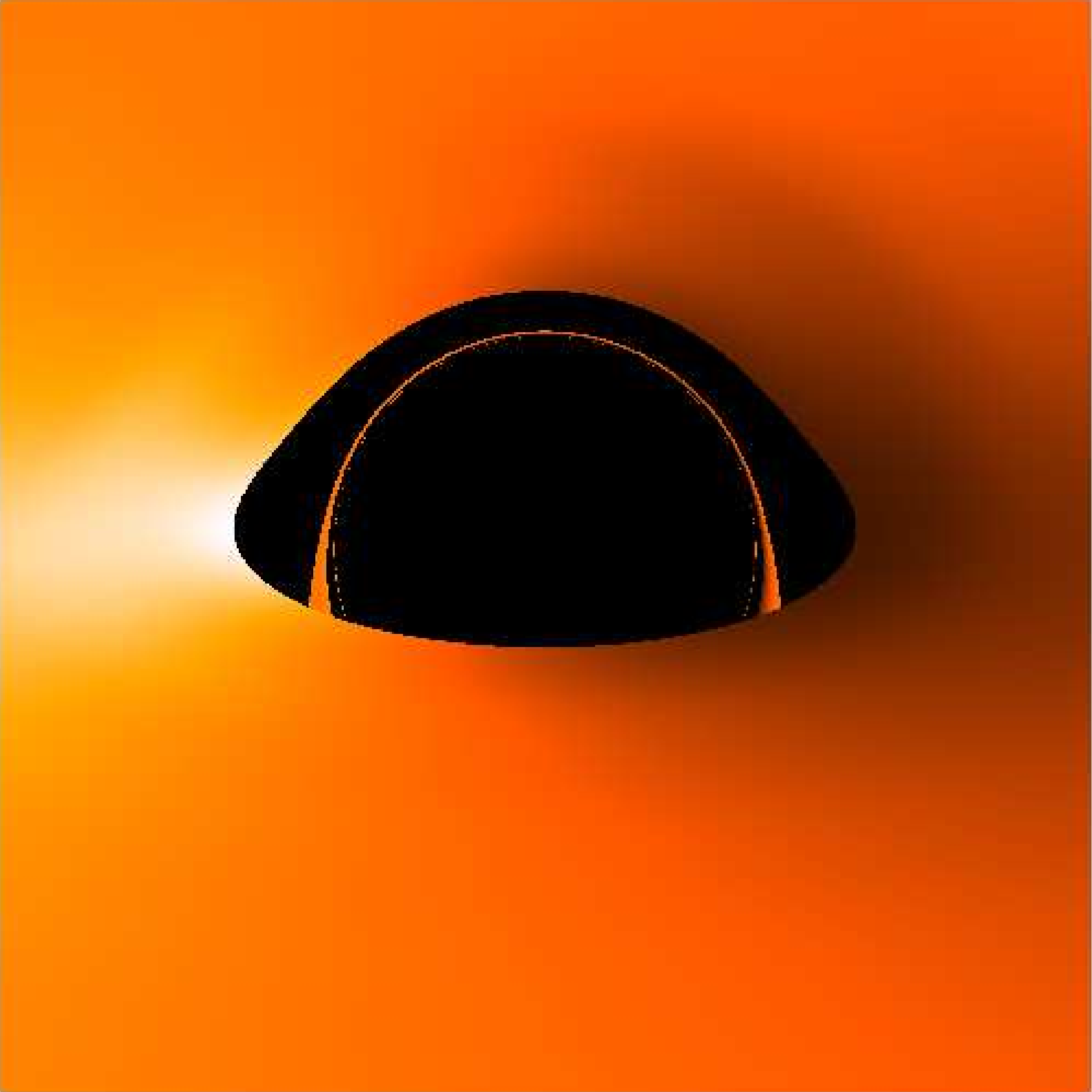} & \includegraphics[width=.3\linewidth]{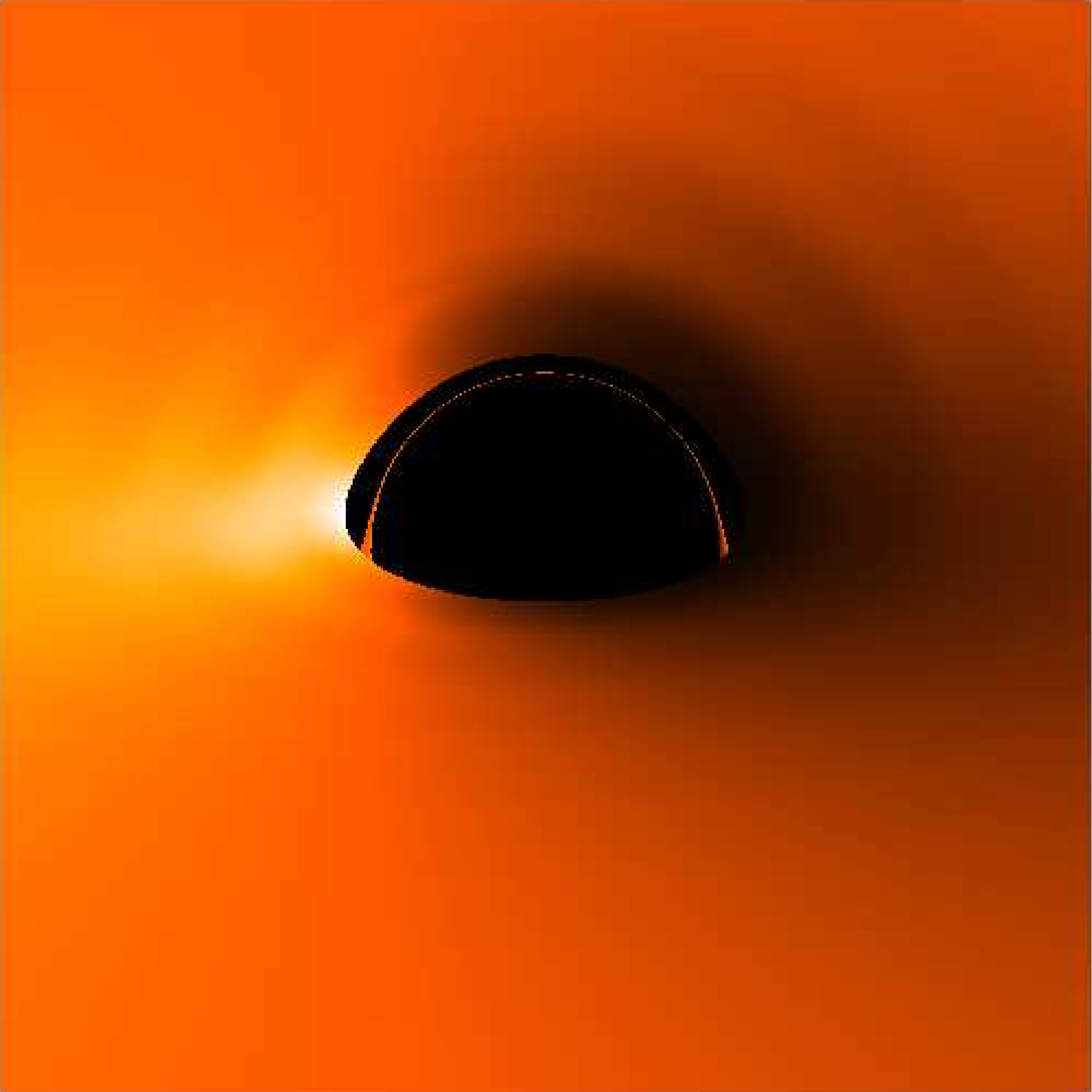}\\
        $\theta=86^\circ$ & 
        \includegraphics[width=.3\linewidth]{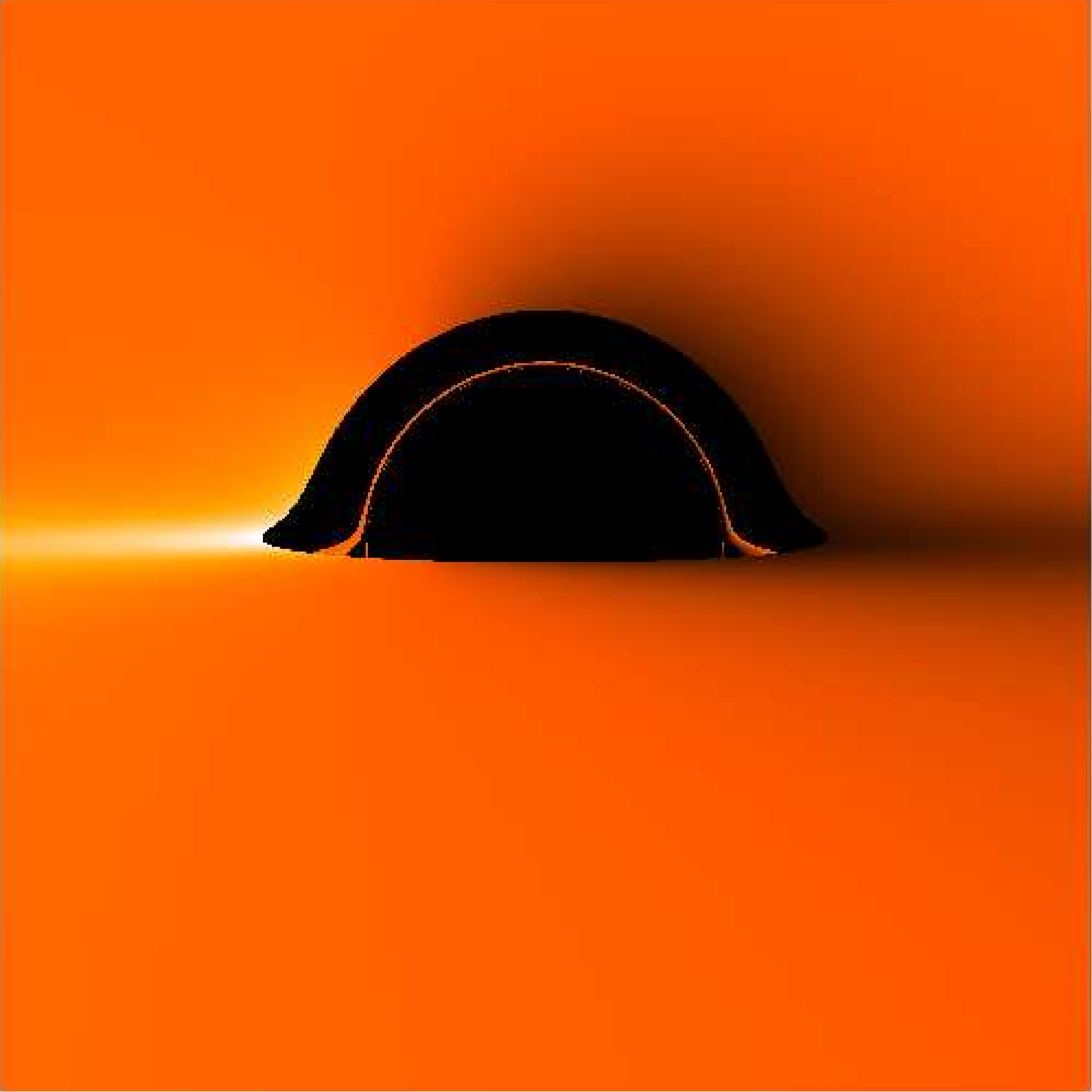} & \includegraphics[width=.3\linewidth]{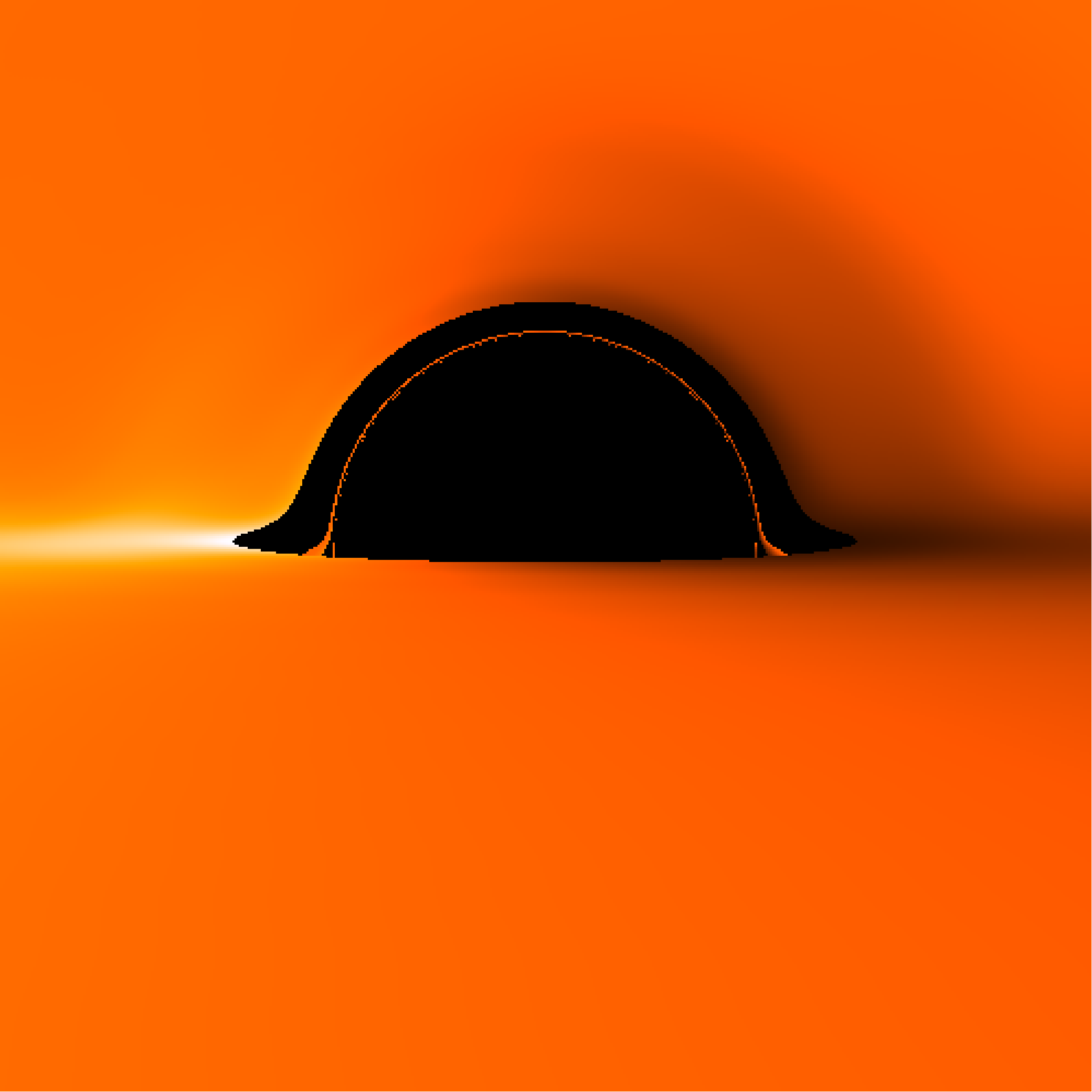} & \includegraphics[width=.3\linewidth]{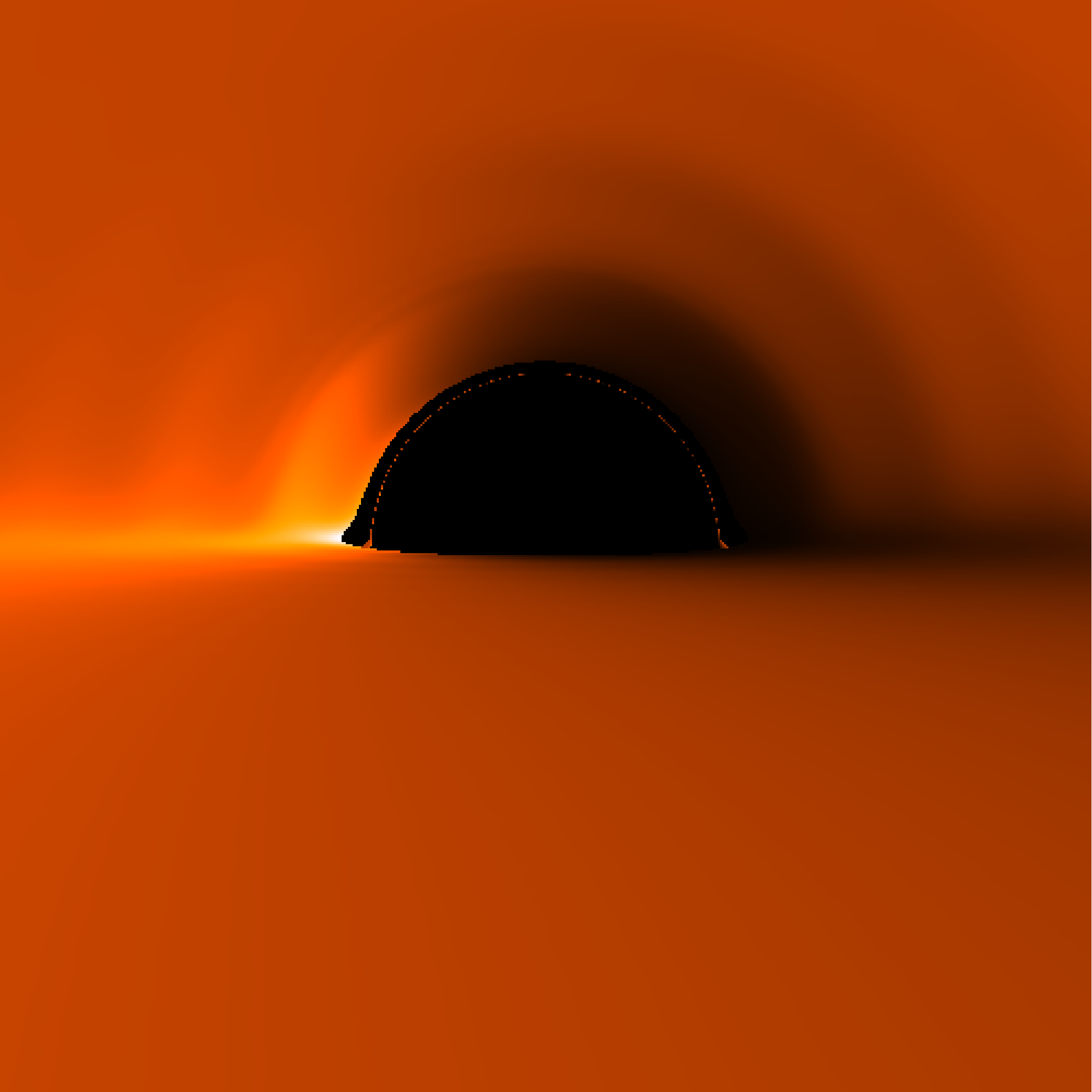}\\
         & Schwarzschild & $R_s=1.06$ & $R_s=0.6$\\
    \end{tabular}
    \caption{Images of geometrically thin and optically thick accretion disks around Schwarzschild and CQBH with $R_s=1.06$ and $R_s=0.6$ for inclination angles $\theta=5^\circ,$ $72^\circ$, $86^\circ$. It is important to notice that despite of the fact that for $R_s\simeq 1.06$ the horizon of the CQBH has value close to the Schwarzschild case the images of the two shadows are considerably different.}
    \label{fig:your_label}
\end{figure*}

We now turn our attention to the shadow image of a CQBH. 
The first computation of the appearance of an accretion disk surrounding a Schwarzschild black hole using the ray-tracing method was done by Luminet in the 1970s~\cite{luminet}. 
To obtain the image of the shadow for the CQBH we use an open-source general relativistic ray-tracing code called \texttt{GYOTO}~\cite{gyoto,Aimar:2023vcs,Paumard_fvincent_Gourgoulhon_LamyFrederic_Torrance_iurso_2024}, which we modify to
simulate the image of a thin infinite accretion disk around the CQBH
and produce images of the gravitationally lensed disk by ray-tracing the photons emitted from the accreting matter towards the observer. 
The \texttt{GYOTO} code integrates geodesics backwards in time, so the initial conditions are given on the observer's screen in terms of position and the incidence angle of the photons. 
Since most photons emitted from the accreting matter do not arrive at the observer, backward ray-tracing deals only with photons that hit the screen thus substantially reducing the number of calculations. 
To integrate the geodesic equations, the code uses the Runge-Kutta algorithm of fourth order \texttt{RK4} with an adaptive step~\cite{numerical}.
The \texttt{GYOTO} code is also capable of producing shadow images with other more realistic models of accretion disks such as Novikov-Thorne~\cite{Novikov:1973kta}, Polish doughnut~\cite{Abramowicz:2011xu}, etc.. 
However for our purposes simulating the shadow of a geometrically thin infinite accretion disk around a CQBH is sufficient to outline the key characteristics that distinguish it from Schwarzschild.

We modified \texttt{GYOTO} to include the CQBH spacetime and for computational complexity reduction, we used the simplest default model of accretion disk called \texttt{‘ThinDisk’}, which describes the geometrically thin accretion disk \cite{Page:1974he,Marck:1995kd}. Geometrically thin refers to the condition $h \ll r$, where $r$ is the radial coordinate on the disk and $h$ is the semi-thickness of the disk. The interior radius of a disk corresponds to $R_{\rm ISCO}$ and the outer radius of the disk is taken at infinity. Also, for simplicity, we consider an optically thick disk, for which only photons emitted from the disk's surface can be observed, i.e. we take $\lambda\ll h$ where $\lambda$ is the mean free path of the photon. 

For the optically thick and geometrically thin disk, according to \cite{fanton1997}, the emitted radiation can be modelled via monochromatic radiation with isotropic specific intensity in the rest frame of the disk. This means that the intensity of emitted radiation $I_{\nu_{em}}$ as a function of the emitted frequency $\nu$ behaves as:
\begin{equation}
    I_{\nu_{em}}\propto\delta(\nu_{\rm em}-\nu_{\rm line})\epsilon(r),
\end{equation}
where $\delta$ is a Dirac distribution, $\nu_{\rm line}$ is the frequency of the monochromatic emission line and $\epsilon(r)$ is the surface emissivity. Here for simplicity we follow the \texttt{GYOTO}'s default model \texttt{‘ThinDisk’} thus taking $\epsilon(r)=1$.
The specific intensity for distant observer is given as:
\begin{equation}
    I_{\nu_{\rm obs}}=g^3I_{\nu_{\rm em}},
\end{equation}
where $\nu_{\rm obs}$ is observed frequency and $g=\nu_{\rm obs}/\nu_{\rm em}$ accounts for the gravitational redshift. The observed flux is then connected to the observed specific intensity via
\begin{equation}
    dF_{\nu_{\rm obs}}=I_{\nu_{\rm obs}}\cos{\alpha}d\Omega,
\end{equation}
where $\alpha$ is the angle between the direction of the incoming photon and the normal to the screen and $d\Omega$ (not to be confused with the line element on the unit sphere) is the infinitesimal solid angle around the direction of incidence of the photon. In the code, to each pixel on the screen corresponds a direction in the sky from which photons can be received. The whole screen is then defined as a collection of pixels covering a given field of view in the sky and the whole screen then covers a finite solid angle $\Omega$.
Then we can define the total observed flux as:
\begin{equation}
F_{\nu_{\rm obs}}=\sum_{i}I_{\nu_{\rm obs}, i}\cos{\alpha_i}\delta\Omega_{i}
\end{equation}
where the subscript `${i}$' corresponds to one pixel, the summation is over the total number of pixels, $\alpha_i$ is the incidence angle for each pixel, $\delta\Omega_{i}$ is the infinitesimal solid angle 
for each pixel, and $I_{\nu_{\rm obs}, i}$ is the observed specific intensity for each pixel.

The simulations of a thin accretion disk around a CQBH are shown in Fig.~\ref{fig:your_label}. 
The figure shows shadow images at three different inclination angles of the line of sight between the screen and the source, i.e. $ \theta = 5^\circ, \ 72^\circ \ \text{and} \ 86^\circ $. Notice that $\theta=0^\circ$ corresponds to the line of sight being perpendicular to the disk and $\theta=90^\circ$ corresponds to the line of sight on the disk's plane. The CQBH shadow images are obtained for two values of the parameter $R_s = 1.06$ and $R_s=0.6$. The first column shows the shadow in the Schwarzschild case for comparison (i.e. $R_s=0$). 
It is immediately noticeable that the CQBH exhibits periodic bright and dim patterns in the accretion disk for both values of $R_s$.  
This alternating bright and dark regions in the accretion disk are a consequence of multiple stable and unstable regions in the effective potential. This behavior is in sharp contrast to the Schwarzschild case.

The second interesting feature is that the inner edge of the accretion disk is much closer to the horizon for the CQBH with $R_s = 0.6$ than the Schwarzschild case. On the other hand, for $R_s = 1.06$, the inner edge is almost at the same distance as the Schwarzschild case. This feature is evident from the variable radius of the ISCO of CQBH which has a sharp dependence on $R_s$ as seen from the Fig.~\ref{fig:risco}.
Therefore, even though for some values of $R_s$ the CQBH and Schwarzschild may have similar values for the location of the horizon or the ISCO radius, the shadow image always looks different from Schwarzschild. This suggest that, depending on the observational capabilities (basically on the power to resolve the bright and dim regions) of the current and future VLBI telescope, it could be possible to distinguish a CQBH from a Schwarzschild black hole via observations.

\section{Discussion}\label{sec6}

Spacetime singularities are one of the most fascinating predictions of GR while, at the same time, being an indicator of the limits of the theory. The main problem with the singularity at the center of black holes is the divergence of curvature and tidal forces that signals a breakdown of the theory.
There exist many attempts to resolve the black hole singularity by introducing suitable corrections to GR in the strong field (see \cite{Malafarina:2017csn} for a recent review). However, a large number of these attempts predict regular black holes (see for example \cite{Malafarina2023}). Such regular black holes, while being non-singular, typically posses a Cauchy horizon, which cause other problems such as mass inflation. 
Some authors claim that singularities are not necessarily a problem since it may be impossible to test the true nature of black hole candidates~\cite{Nakao2024}.
The CQBH considered here replaces the singular center with a finite quantum core which exhibits a classical behavior and has an integrable, i.e. tame, singularity, without a Cauchy horizon. This makes it an appealing hypothetical candidate for astrophysical compact objects. Moreover, we have shown that the `classicalization' of the quantum corrections which produce the object's core has macroscopic effects that produce distinct features in the observable properties of the CQBH's accretion disk.
Such effects may be in principle observed thus allowing to test and eventually rule out the model.

For example, the bright and dark ring signatures in the CQBH's shadow lead to the possibility of distinguishing it from the Schwarzschild metric by astrophysical observations. 
The periodicity of $2 \pi R_s$ in the function ${\rm Si}(r/R_s)$ would induce a corresponding periodicity in the separation $d$ between nearby rings in the accretion disk. 
In principle this value can be related to the distance between bright and dark rings in the image of the accretion disk as obtained by a telescope thus allowing to put constraints on the allowed values of $R_s$ from shadow and lensing observations.

\section*{Acknowledgement}
The authors would like to thank Roberto Casadio and Andrea Giusti for useful comments and discussion.
The authors acknowledge support from Nazarbayev University Faculty Development Competitive Research Grant No. 11022021FD2926.

\bibliography{ref}

\end{document}